\documentclass[twocolumn,aps,prl,showpacs,superscriptaddress,twocolumn]{revtex4-1}
\usepackage{lmodern}

\usepackage[latin9]{inputenc}
\usepackage{amsmath}
\usepackage{amssymb}
\usepackage{graphicx}
\usepackage{epstopdf}
\usepackage{color}
\usepackage{enumerate}
\usepackage{ulem}

\begin{document}
\title{A Direct Determination of the Temperature of Overheated Electrons in an Insulator}

		\author{T. Levinson}
		\email{Levinsontal@Gmail.com.; Corresponding author}
		\affiliation{Department of Condensed Matter Physics, The Weizmann Institute of Science, Rehovot 76100, Israel.}
		
		\author{A. Doron}
		\affiliation{Department of Condensed Matter Physics, The Weizmann Institute of Science, Rehovot 76100, Israel.}
		
		\author{I. Tamir}
		\affiliation{Department of Condensed Matter Physics, The Weizmann Institute of Science, Rehovot 76100, Israel.}
		
		\author{G. C. Tewari}
		\affiliation{Department of Condensed Matter Physics, The Weizmann Institute of Science, Rehovot 76100, Israel.}
	 
		\author{D. Shahar}
		\affiliation{Department of Condensed Matter Physics, The Weizmann Institute of Science, Rehovot 76100, Israel.}

	\begin{abstract}
		Highly disordered superconductors, in the magnetic-field-driven insulating state, can show discontinuous current-voltage characteristics. Electron overheating has been shown to give a consistent description of this behavior, but there are other, more exotic, explanations including a novel, "superinsulating" state and a many-body localized state. We present AC-DC crossed-measurements, in which the application of a DC voltage is applied along our sample, while a small AC voltage is applied in the transverse direction. We varied the DC voltage and observed a simultaneous discontinuity in both AC and DC currents. We show that the inferred electron-temperature in the transverse measurement matches that in the longitudinal one, strongly supporting electron overheating as the source of observed current-voltage characteristics. Our measurement technique may be applicable as a method of probing electron overheating in many other physical systems, which show discontinuous or non-linear current-voltage characteristics.
	\end{abstract}
\maketitle

Highly disordered superconductors can undergo a transition to an insulating state. This superconductor to insulator transition (SIT) can be driven by several parameters such as disorder strength, thickness or magnetic field ($B$)\cite{HebardPrl,goldmanpt51,sondhirmp,physupekhi}. While studying the $B$-driven insulating state in amorphous Indium Oxide (a:InO) thin films, Sambandamurthy et al. discovered that, at low temperature ($T$), discontinuities appear in the current-voltage characteristics ($I-V's$)\cite{murthyprl}. Similar findings were later seen in disordered Titanium Nitride thin films\cite{BaturinaPRL}, where they have been interpreted as evidence for a novel insulating state, termed a superinsulator\cite{vinokurnat}.

More recently, Altshuler et al.\cite{borisprl} argued that the discontinuous $I-V's$ can be accounted for by electron overheating. Their theory is based on the assumptions that the electrons interact weakly with the phonons but strongly with each other, thus leading to the possibility that they will have their own, well defined $T$ ($T_{el}$), which can be very different from that of the phonons ($T_{ph}$). Additionally, they assumed that Ohm's law holds for the entire voltage ($V$) range of the measurements, i.e. $V=I\cdot R(T_{el})$, where $I$ is the current and $R$ is the resistance.  In a steady state, they could obtain $T_{el}$ from a heat-balance eq:
	\begin{equation}
	\frac{V^2}{R(T_{el}(V))}=\Gamma\Omega*(T_{el}^{\beta}-T_{ph}^{\beta}),
	\label{e1}
	\end{equation}
Where $\Omega$ is the sample volume, $\Gamma$ is the electron-phonon coupling coefficient\cite{borisprl,maozprl} and $\beta$ is a material-dependent constant. 

By numerically solving eq.\ref{e1} for the experimentally relevant parameters, Altshuler et al.\cite{borisprl} found that, below a certain $T_{ph}$, $T_{el}(V)$ develops a bi-stable region, i.e., for a certain range of $V$ eq.\ref{e1} can have two stable solutions: A low $T_{el}$ solution, in which $T_{el}\approx T_{ph}$, resulting in high $R$, and a high $T_{el}$ solution, in which $T_{el}$ can be much higher than $T_{ph}$, with much lower $R$. At equilibrium, ($V=0$) the system is in the low $T_{el}$ solution. As $V$ is increased above a threshold value, the system enters the bi-stable region where it can spontaneously jump to the high $T_{el}$ solution, resulting in a discontinuous jump in $I$. Alongside ref.\cite{borisprl}, Ovadia et al. conducted a detailed experimental study of the $I-V's$, showing that they are consistent with the overheated electron framework\cite{maozprl}. 

If this framework properly describes the physics behind observed $I-V's$, an interesting scenario emerges. At low $T$ the application of a $V$ can result in an analogue of the liquid to gas phase transition, but under non-equilibrium conditions\cite{AdamPRL}. The electronic system can be driven far from equilibrium, offering an experimental tool to study the nature of quantum systems under such conditions.

Despite the consistency shown by the experimental results of Ovadia et al., a direct demonstration that electron overheating is behind the reported $I-V's$ via a direct measurement of $T_{el}$, is still lacking. This demonstration is essential because there are other theoretical approaches that offer a distinctly different view of the discontinuous $I-V's$ in our, and in others', systems\cite{murthyprl,maozprl,BaturinaPRL,PhysRevLett.74.3237,PhysRevB.90.195443,PhysRevB.57.R6842,Pilling1996652,PhysRevB.53.973}. One such theory is that the low $I$ branch of the experimental $I-V's$ is evidence of a "superinsulating" state which is destroyed at a critical $V$\cite{vinokurnat}, dual to the critical $I$ in a superconductor. Another competing theory is that the observed $I-V's$ are a manifestation of a novel many-body localized state, as explained in ref.\cite{baskoprb}. A third possible explanation is that application of an electric field ($E$) tilts the random potential created by disorder until, at a threshold $E$ value, a conduction channel connecting the two ends of the sample forms. This model was treated in ref.\cite{PhysRevLett.71.3198} in the context of metallic islands in a disordered potential. It was used in order to explain discontinuous and non-linear $I-V's$ observed in various systems\cite{PhysRevB.53.973,PhysRevLett.74.3237,PhysRevB.57.R6842,PhysRevB.90.195443}, including insulating films\cite{PhysRevB.53.973}. 

We considered electronic noise measurements as a possible method to directly measure $T_{el}$. These measurements, an obvious option because equilibrium noise is a commonly used thermometer, proved unfeasible. The combination of low $T_{el}$ ($\approx50$ mK), high $R$ (typically $>10^8\Omega$) and the need to flow a DC $I$ while conducting the measurement, is yet too challenging experimentally. 

In this letter we have taken a different approach: According to ref.\cite{borisprl}, $R$ is determined by $T_{el}$ that, in turn, is determined by the power input into the system via Joule heating. Under the assumption that Joule heating is uniform, $T_{el}$ is also expected to be uniform throughout the system. As a result, $T_{el}$, as inferred by a measurement of $R$ in a given orientation, should be the same regardless of the direction in which $V$ is applied. The crux of our method was to sweep $V$ and infer $T_{el}$ from two independent measurements of $R$, one in the direction parallel to $V$ and another in a perpendicular direction.

It is important to note that our method cannot directly distinguish between the scenario described above, where $T_{el}$ differs from $T_{ph}$, which in turn remains equal to the $T$ of the external bath ($T_B$), and a scenario where the entire sample decouples from the external bath, i.e. $T_{el}=T_{ph}\neq T_B$. In order to distinguish between these two scenarios we conducted a test experiment, whose details we describe in the supplementary materials section of this paper, which clearly indicates that the first scenario is the relevant one in our system.

	\begin{figure} [t]
		\includegraphics[width=1\linewidth]{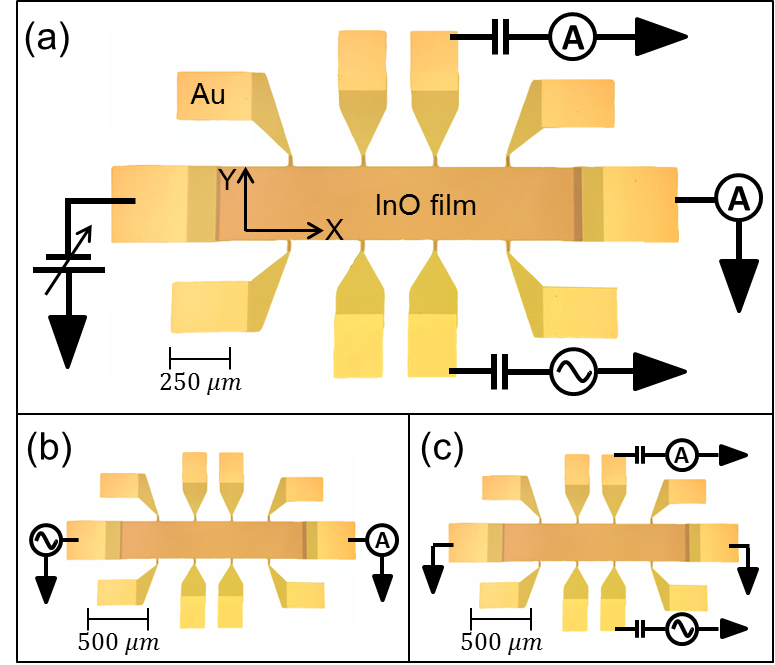} \caption{\label{Setup}\textbf{Our experimental setup}. \textbf{a}. shows the setup we used to measure $I_{X,Y}(V_X)$ (for more information see supplementary). The electronic circuit shown schematically was comprised of two trans-impedance amplifiers, a Lock-in amplifier, a DC $V$-source and two $10\mu$F capacitors used as DC-blocks. \textbf{b}. shows the setup we used to measure $\left. R_X\right| _{V=0}(T)$. \textbf{c}. shows the AC-equivalent electronic circuit of the full experimental setup, which is the setup we used to measure $\left. R_Y\right| _{V=0}(T)$.
		}
	\end{figure}
	
	\begin{figure} [t]
		\includegraphics[width=1\linewidth]{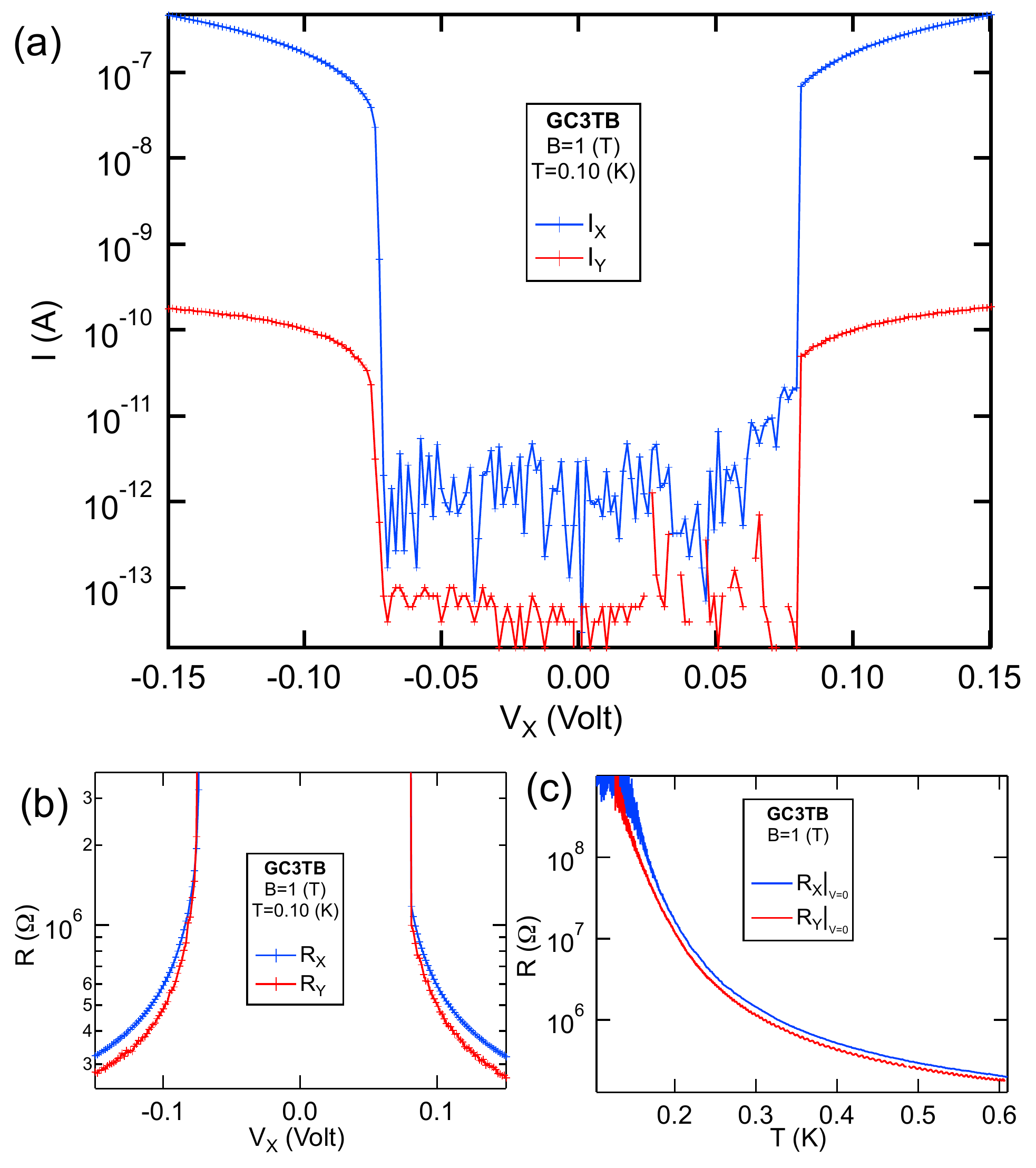} 
		\caption{\textbf{a}. $I_X$ and $I_Y$ vs $V_X$. While $V_X$ was swept, a constant $V_Y=50$ mV (RMS) was applied and $I_X,I_Y$ were simultaneously measured. \textbf{b}. $R_X$ and $R_Y$ vs $V_X$, calculated from the data shown in sub-fig. a. \textbf{c}. $\left. R_X\right| _{V=0}(T)$ and $\left. R_Y\right| _{V=0}(T)$ vs $T$. All sub-figures are plotted on a semi-log scale. The data were measured at $B=1$ T, the data in sub-figs. a and b were taken at $T=100$ mK. \label{IV,RV,RT}}
	\end{figure}
The results presented in this work were obtained from a $1500\mu m$ long, $300\mu m$ wide, 10 contact Hall-bar sample of $30nm$ thick a:InO (see fig.\ref{Setup}).The critical $B$ ($B_c$) at which our sample showed a SIT was $0.009$ T. A DC $V$ was applied in the longitudinal (X) direction while a small, $50\mu V_{RMS}$ AC, $V$ was applied in the transverse (Y) direction (see supplementary information for more details). The longitudinal $V$ ($V_X$) was swept from $-150$ to $150$ mV, while the transverse $V$ ($V_Y$) was held constant. During this, the longitudinal DC $I$ ($I_X$) and the transverse AC $I$ ($I_Y$) were measured. 

$I_X$ and $I_Y$ vs $V_X$ are shown in fig.\ref{IV,RV,RT}a. At low $V_X$ both $I_X$ and $I_Y$ are within the noise. As $V_X$ is increased, at $V_X\approx80$ mV, $I_X$ and $I_Y$ each attain a value that appears abruptly, well above the noise, and grows progressively as $V_X$ is increased further. Upon reducing $V_X$ (Going from $-150$ to $0$ mV), $I_X$ and $I_Y$ decrease gradually, until rapidly dropping at $V_X\approx72.5mV$. 

Following the assumptions stated above, we calculated the longitudinal and transverse $R's$ ($R_X$ and $R_Y$) by applying Ohm's law: $R_{X,Y}=V_{X,Y}/I_{X,Y}$. In fig.\ref{IV,RV,RT}b, $R_X$ and $R_Y$ are plotted against $V_X$, showing the same abrupt escape from the noise as seen for the $I's$. The maximum $R$ that we could measure, given our noise level in this setup, was $\approx10^8$ $\Omega$. 

In order to infer a value of $T_{el}$ from each of the data points, we used the zero-bias $R(T)$ of the sample($\left. R \right|_{V=0}(T)$), where $T$ refers to the cryostat temperature. We conducted a two-terminal measurement of $\left. R \right|_{V=0}(T)$, while sweeping $T$ slowly (in order to ensure $T_{ph}=T$). At low $V$ the $I-V's$ were linear, as was verified by $I-V$ measurements (see supplementary), indicating that $T_{el}=T_{ph}$. Thus by measuring in the linear range we ensured  $\left. R \right|_{V=0}(T)=R(T_{el})$.

We measured $\left. R \right|_{V=0}(T)$ in two different setups. The first was the setup, shown in fig.\ref{Setup}b, which we used to measure $\left. R_X \right|_{V=0}(T)$, and the second was the AC-equivalent of our experimental setup, shown in fig.\ref{Setup}c, which we used to measure $\left. R_Y \right|_{V=0}(T)$. In both cases we used an AC $V$ of $50\mu V_{RMS}$ which is well within the linear $I-V$ range. $\left. R_X \right|_{V=0}$ and $\left. R_Y \right|_{V=0}$ vs $T$ are shown in fig.\ref{IV,RV,RT}b.

\begin{figure} [t]
		\includegraphics[width=1\linewidth]{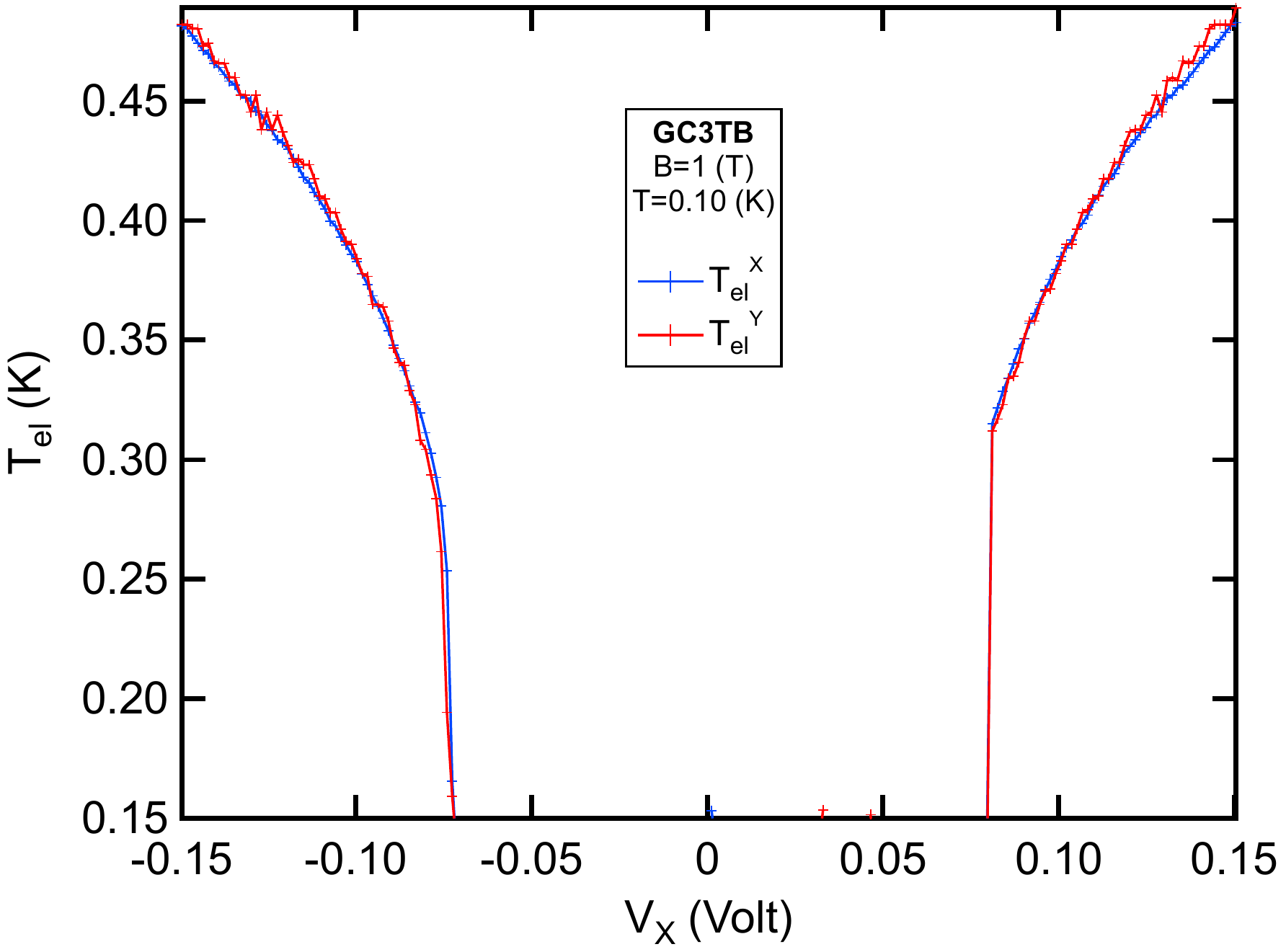} \caption{\label{Tel_V_Example}$T_{el}^X$ and $T_{el}^Y$ vs $V_X$, plotted on a linear scale. $T_{el}^{X,Y}$ were calculated from $R_{X,Y}$ and $\left. R_{X,Y}\right| _{V=0}(T)$ plotted in fig.\ref{IV,RV,RT}. Our experimental error did not allow us to infer a $T_{el}$ value of below 150 mK in a reliable way, thus no data points appear below this value.}
	\end{figure}

We inverted $\left. R_{X,Y}\right|_{V=0}(T)$, attained $T_{el}^{X,Y}\equiv T_{el}(R_{X,Y})$ and inferred a $T_{el}$ value from each data point of $R_{X,Y}(V_X)$.  In fig.\ref{Tel_V_Example} we plotted $T_{el}^X$ and $T_{el}^Y$ vs $V_X$. Our experimental error in $R_{X,Y}$ allowed us to measure $T_{el}$ down to $~150$ mK. The agreement between $T_{el}^X$ and $T_{el}^Y$ is near-perfect.

\begin{figure} [t]
		\includegraphics[width=1\linewidth]{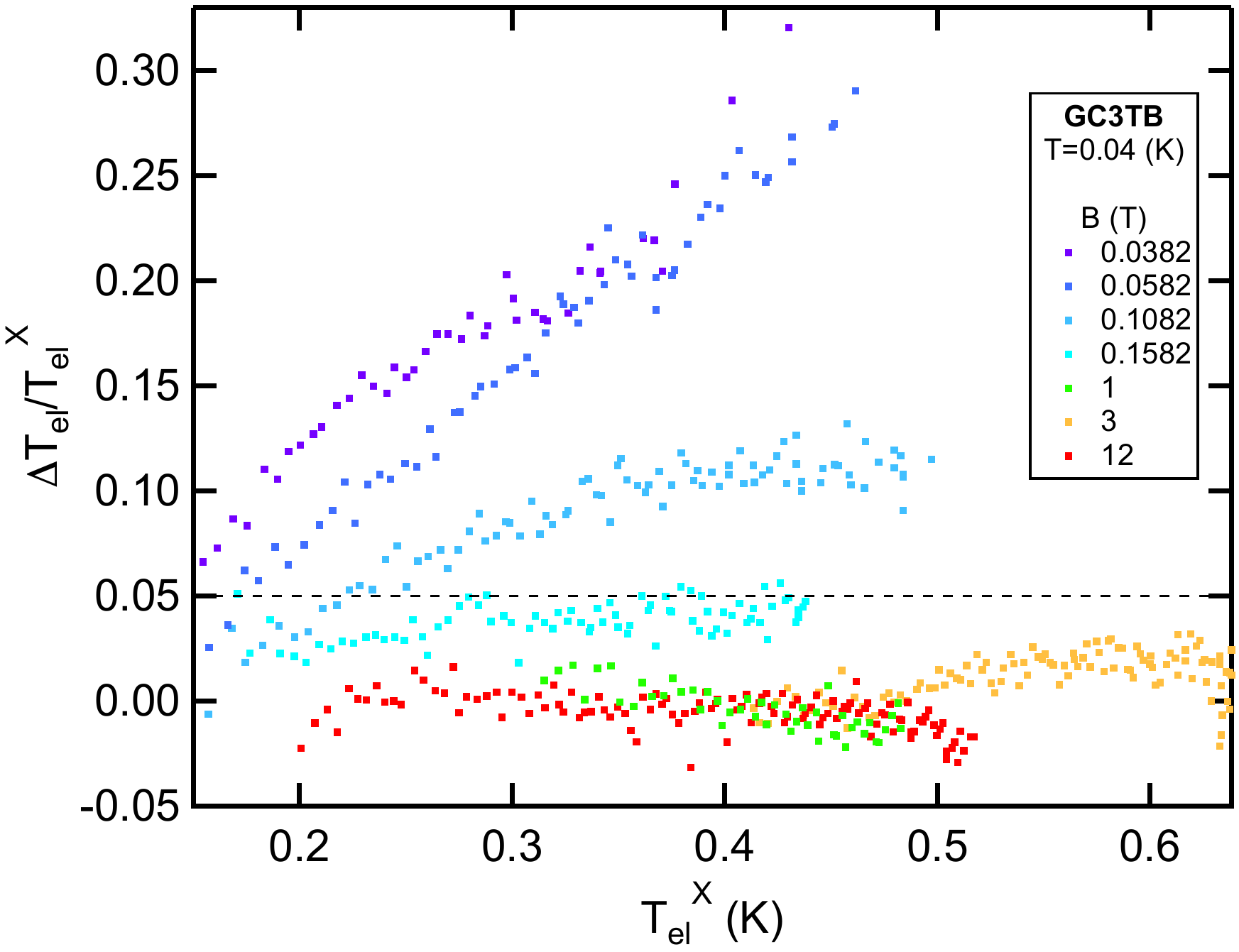} \caption{\label{DelTel}$\frac{\Delta T_{el}}{T_{el}^X}$ vs $T_{el}^X$, plotted on a linear scale. The dashed line represents $\frac{\Delta T_{el}}{T_{el}^X}=5\%$. The color-scale represent the $B$ corresponding to each data set. All data were calculated from measurements conducted at $T=40$ mK and from $\left. R_{X,Y}\right| _{V=0}(T)$ similar to those plotted in fig.\ref{IV,RV,RT}.}
	\end{figure}

We next turned our focus to a $B$-dependent study of the $I-V's$ using our new technique. We defined $\Delta T_{el}\equiv T_{el}^X-T_{el}^Y$. In fig.\ref{DelTel} we plotted $\frac{\Delta T_{el}}{T_{el}^X}$ vs $T_{el}^X$ for various $B's$ ranging from 0.0382 up to 12 T. For sake of clarity we only displayed a portion of the $B's$ that were studied, which represents the trend observed for all $B's$ examined. For $B>0.15$ T, all measured $T_{el}^X$ were within $5\%$ of $T_{el}^Y$ i.e. below the dashed line in fig.\ref{DelTel}. For $B<0.15$ T, as $B$ approached $B_c$, $T_{el}^X$ became systematically larger than $T_{el}^Y$. Both noise and systematic error in $T_{el}^{X,Y}$ grew significantly as $B$ approached $B_c$. This is due to $R(T_{el})$ becoming a progressively slow-varying function of $T_{el}$, thus inverting it caused large uncertainty. Nevertheless, our measurements point to a possible deviation from the overheated electron picture as $B_c$ is approached. 

The main conclusion of our work stems from the near-perfect agreement between $T_{el}^X$ and $T_{el}^Y$, which strongly supports the theory of electron overheating as the cause of observed $I-V's$ in the $B$-driven insulating state displayed by disordered films. We identified a possible deviation from the overheated electron framework in the vicinity of $B_c$, requiring further investigation. We demonstrated a novel, simple technique to measure electronic temperature of systems which exhibit non-linear $I-V's$. Many condensed matter systems such as Anderson-Mott insulators\cite{PhysRevB.53.973}, semiconductor Quantum dot arrays\cite{PhysRevLett.74.3237,PhysRevB.90.195443}, metallic island arrays\cite{PhysRevB.57.R6842} and transport through constrictions\cite{Pilling1996652} exhibit similar $I-V's$. Applying our technique to these systems may serve to enhance our understanding of them, and show whether electron overheating is involved in the non-linear $I-V's$ observed in these systems. 

	\section*{Acknowledgments}
	We would like to acknowledge B.L. Altshuler for suggesting this experiment and V.E. Kravtsov for fruitful discussions on the subject. This work was supported by the Israeli Science Foundation and the United States-Israel Bi-national Science Foundation.
	
	

\par


\bibliographystyle{apsrev}

\begin{thebibliography}{17}
	\expandafter\ifx\csname natexlab\endcsname\relax\def\natexlab#1{#1}\fi
	\expandafter\ifx\csname bibnamefont\endcsname\relax
	\def\bibnamefont#1{#1}\fi
	\expandafter\ifx\csname bibfnamefont\endcsname\relax
	\def\bibfnamefont#1{#1}\fi
	\expandafter\ifx\csname citenamefont\endcsname\relax
	\def\citenamefont#1{#1}\fi
	\expandafter\ifx\csname url\endcsname\relax
	\def\url#1{\texttt{#1}}\fi
	\expandafter\ifx\csname urlprefix\endcsname\relax\def\urlprefix{URL }\fi
	\providecommand{\bibinfo}[2]{#2}
	\providecommand{\eprint}[2][]{\url{#2}}
	
	\bibitem[{\citenamefont{Hebard and Paalanen}(1990)}]{HebardPrl}
	\bibinfo{author}{\bibfnamefont{A.~F.} \bibnamefont{Hebard}} \bibnamefont{and}
	\bibinfo{author}{\bibfnamefont{M.~A.} \bibnamefont{Paalanen}},
	\bibinfo{journal}{Phys. Rev. Lett.} \textbf{\bibinfo{volume}{65}},
	\bibinfo{pages}{927} (\bibinfo{year}{1990}).
	
	\bibitem[{\citenamefont{Goldman and Markovic}(1998)}]{goldmanpt51}
	\bibinfo{author}{\bibfnamefont{A.~M.} \bibnamefont{Goldman}} \bibnamefont{and}
	\bibinfo{author}{\bibfnamefont{N.}~\bibnamefont{Markovic}},
	\bibinfo{journal}{Phys. Today} \textbf{\bibinfo{volume}{51}},
	\bibinfo{pages}{39} (\bibinfo{year}{1998}).
	
	\bibitem[{\citenamefont{Sondhi et~al.}(1997)\citenamefont{Sondhi, Girvin,
			Carini, and Shahar}}]{sondhirmp}
	\bibinfo{author}{\bibfnamefont{S.~L.} \bibnamefont{Sondhi}},
	\bibinfo{author}{\bibfnamefont{S.~M.} \bibnamefont{Girvin}},
	\bibinfo{author}{\bibfnamefont{J.~P.} \bibnamefont{Carini}},
	\bibnamefont{and} \bibinfo{author}{\bibfnamefont{D.}~\bibnamefont{Shahar}},
	\bibinfo{journal}{Rev. Mod. Phys.} \textbf{\bibinfo{volume}{69}},
	\bibinfo{pages}{315} (\bibinfo{year}{1997}).
	
	\bibitem[{\citenamefont{Gantmakher and Dolgopolov}(2010)}]{physupekhi}
	\bibinfo{author}{\bibfnamefont{V.~F.} \bibnamefont{Gantmakher}}
	\bibnamefont{and} \bibinfo{author}{\bibfnamefont{V.~T.}
		\bibnamefont{Dolgopolov}}, \bibinfo{journal}{Phys.-Usp.}
	\textbf{\bibinfo{volume}{53}}, \bibinfo{pages}{1} (\bibinfo{year}{2010}).
	
	\bibitem[{\citenamefont{Sambandamurthy
			et~al.}(2005)\citenamefont{Sambandamurthy, Engel, Johansson, Peled, and
			Shahar}}]{murthyprl}
	\bibinfo{author}{\bibfnamefont{G.}~\bibnamefont{Sambandamurthy}},
	\bibinfo{author}{\bibfnamefont{L.~W.} \bibnamefont{Engel}},
	\bibinfo{author}{\bibfnamefont{A.}~\bibnamefont{Johansson}},
	\bibinfo{author}{\bibfnamefont{E.}~\bibnamefont{Peled}}, \bibnamefont{and}
	\bibinfo{author}{\bibfnamefont{D.}~\bibnamefont{Shahar}},
	\bibinfo{journal}{Phys. Rev. Lett.} \textbf{\bibinfo{volume}{94}},
	\bibinfo{pages}{017003} (\bibinfo{year}{2005}).
	
	\bibitem[{\citenamefont{Baturina et~al.}(2007)\citenamefont{Baturina, Mironov,
			Vinokur, Baklanov, and Strunk}}]{BaturinaPRL}
	\bibinfo{author}{\bibfnamefont{T.~I.} \bibnamefont{Baturina}},
	\bibinfo{author}{\bibfnamefont{A.~Y.} \bibnamefont{Mironov}},
	\bibinfo{author}{\bibfnamefont{V.~M.} \bibnamefont{Vinokur}},
	\bibinfo{author}{\bibfnamefont{M.~R.} \bibnamefont{Baklanov}},
	\bibnamefont{and} \bibinfo{author}{\bibfnamefont{C.}~\bibnamefont{Strunk}},
	\bibinfo{journal}{Phys. Rev. Lett.} \textbf{\bibinfo{volume}{99}},
	\bibinfo{pages}{257003} (\bibinfo{year}{2007}).
	
	\bibitem[{\citenamefont{Vinokur et~al.}(2008)\citenamefont{Vinokur, Baturina,
			Fistul, Mironov, Baklanov, and Strunk}}]{vinokurnat}
	\bibinfo{author}{\bibfnamefont{V.~M.} \bibnamefont{Vinokur}},
	\bibinfo{author}{\bibfnamefont{T.~I.} \bibnamefont{Baturina}},
	\bibinfo{author}{\bibfnamefont{M.~V.} \bibnamefont{Fistul}},
	\bibinfo{author}{\bibfnamefont{A.~Y.} \bibnamefont{Mironov}},
	\bibinfo{author}{\bibfnamefont{M.~R.} \bibnamefont{Baklanov}},
	\bibnamefont{and} \bibinfo{author}{\bibfnamefont{C.}~\bibnamefont{Strunk}},
	\bibinfo{journal}{Nature} \textbf{\bibinfo{volume}{452}},
	\bibinfo{pages}{613} (\bibinfo{year}{2008}).
	
	\bibitem[{\citenamefont{Altshuler et~al.}(2009)\citenamefont{Altshuler,
			Kravtsov, Lerner, and Aleiner}}]{borisprl}
	\bibinfo{author}{\bibfnamefont{B.~L.} \bibnamefont{Altshuler}},
	\bibinfo{author}{\bibfnamefont{V.~E.} \bibnamefont{Kravtsov}},
	\bibinfo{author}{\bibfnamefont{I.~V.} \bibnamefont{Lerner}},
	\bibnamefont{and} \bibinfo{author}{\bibfnamefont{I.~L.}
		\bibnamefont{Aleiner}}, \bibinfo{journal}{Phys. Rev. Lett.}
	\textbf{\bibinfo{volume}{102}}, \bibinfo{pages}{176803}
	(\bibinfo{year}{2009}).
	
	\bibitem[{\citenamefont{Ovadia et~al.}(2009)\citenamefont{Ovadia, Sacepe, and
			Shahar}}]{maozprl}
	\bibinfo{author}{\bibfnamefont{M.}~\bibnamefont{Ovadia}},
	\bibinfo{author}{\bibfnamefont{B.}~\bibnamefont{Sacepe}}, \bibnamefont{and}
	\bibinfo{author}{\bibfnamefont{D.}~\bibnamefont{Shahar}},
	\bibinfo{journal}{Phys. Rev. Lett.} \textbf{\bibinfo{volume}{102}},
	\bibinfo{pages}{176802} (\bibinfo{year}{2009}).
	
	\bibitem[{\citenamefont{Doron et~al.}(2016)\citenamefont{Doron, Tamir, Mitra,
			Zeltzer, Ovadia, and Shahar}}]{AdamPRL}
	\bibinfo{author}{\bibfnamefont{A.}~\bibnamefont{Doron}},
	\bibinfo{author}{\bibfnamefont{I.}~\bibnamefont{Tamir}},
	\bibinfo{author}{\bibfnamefont{S.}~\bibnamefont{Mitra}},
	\bibinfo{author}{\bibfnamefont{G.}~\bibnamefont{Zeltzer}},
	\bibinfo{author}{\bibfnamefont{M.}~\bibnamefont{Ovadia}}, \bibnamefont{and}
	\bibinfo{author}{\bibfnamefont{D.}~\bibnamefont{Shahar}},
	\bibinfo{journal}{Phys. Rev. Lett.} \textbf{\bibinfo{volume}{116}},
	\bibinfo{pages}{057001} (\bibinfo{year}{2016}).
	
	\bibitem[{\citenamefont{Duruoz et~al.}(1995)\citenamefont{Duruoz, Clarke,
			Marcus, and Harris}}]{PhysRevLett.74.3237}
	\bibinfo{author}{\bibfnamefont{C.~I.} \bibnamefont{Duruoz}},
	\bibinfo{author}{\bibfnamefont{R.~M.} \bibnamefont{Clarke}},
	\bibinfo{author}{\bibfnamefont{C.~M.} \bibnamefont{Marcus}},
	\bibnamefont{and} \bibinfo{author}{\bibfnamefont{J.~S.} \bibnamefont{Harris},
		\bibfnamefont{Jr.}}, \bibinfo{journal}{Phys. Rev. Lett.}
	\textbf{\bibinfo{volume}{74}}, \bibinfo{pages}{3237} (\bibinfo{year}{1995}).
	
	\bibitem[{\citenamefont{Staley et~al.}(2014)\citenamefont{Staley, Ray, Kastner,
			Hanson, and Gossard}}]{PhysRevB.90.195443}
	\bibinfo{author}{\bibfnamefont{N.~E.} \bibnamefont{Staley}},
	\bibinfo{author}{\bibfnamefont{N.}~\bibnamefont{Ray}},
	\bibinfo{author}{\bibfnamefont{M.~A.} \bibnamefont{Kastner}},
	\bibinfo{author}{\bibfnamefont{M.~P.} \bibnamefont{Hanson}},
	\bibnamefont{and} \bibinfo{author}{\bibfnamefont{A.~C.}
		\bibnamefont{Gossard}}, \bibinfo{journal}{Phys. Rev. B}
	\textbf{\bibinfo{volume}{90}}, \bibinfo{pages}{195443}
	(\bibinfo{year}{2014}).
	
	\bibitem[{\citenamefont{Kurdak et~al.}(1998)\citenamefont{Kurdak, Rimberg, Ho,
			and Clarke}}]{PhysRevB.57.R6842}
	\bibinfo{author}{\bibfnamefont{C.}~\bibnamefont{Kurdak}},
	\bibinfo{author}{\bibfnamefont{A.~J.} \bibnamefont{Rimberg}},
	\bibinfo{author}{\bibfnamefont{T.~R.} \bibnamefont{Ho}}, \bibnamefont{and}
	\bibinfo{author}{\bibfnamefont{J.}~\bibnamefont{Clarke}},
	\bibinfo{journal}{Phys. Rev. B} \textbf{\bibinfo{volume}{57}},
	\bibinfo{pages}{R6842} (\bibinfo{year}{1998}).
	
	\bibitem[{\citenamefont{Pilling et~al.}(1996)\citenamefont{Pilling, Cobden,
			McEuen, Duruoz, and Jr.}}]{Pilling1996652}
	\bibinfo{author}{\bibfnamefont{G.}~\bibnamefont{Pilling}},
	\bibinfo{author}{\bibfnamefont{D.}~\bibnamefont{Cobden}},
	\bibinfo{author}{\bibfnamefont{P.}~\bibnamefont{McEuen}},
	\bibinfo{author}{\bibfnamefont{C.}~\bibnamefont{Duruoz}}, \bibnamefont{and}
	\bibinfo{author}{\bibfnamefont{J.~H.} \bibnamefont{Jr.}},
	\bibinfo{journal}{Surface Science} \textbf{\bibinfo{volume}{361}},
	\bibinfo{pages}{652 } (\bibinfo{year}{1996}).
	
	\bibitem[{\citenamefont{Ladieu et~al.}(1996)\citenamefont{Ladieu, Sanquer, and
			Bouchaud}}]{PhysRevB.53.973}
	\bibinfo{author}{\bibfnamefont{F.}~\bibnamefont{Ladieu}},
	\bibinfo{author}{\bibfnamefont{M.}~\bibnamefont{Sanquer}}, \bibnamefont{and}
	\bibinfo{author}{\bibfnamefont{J.~P.} \bibnamefont{Bouchaud}},
	\bibinfo{journal}{Phys. Rev. B} \textbf{\bibinfo{volume}{53}},
	\bibinfo{pages}{973} (\bibinfo{year}{1996}).
	
	\bibitem[{\citenamefont{Basko et~al.}(2007)\citenamefont{Basko, Aleiner, and
			Altshuler}}]{baskoprb}
	\bibinfo{author}{\bibfnamefont{D.~M.} \bibnamefont{Basko}},
	\bibinfo{author}{\bibfnamefont{I.~L.} \bibnamefont{Aleiner}},
	\bibnamefont{and} \bibinfo{author}{\bibfnamefont{B.~L.}
		\bibnamefont{Altshuler}}, \bibinfo{journal}{Phys. Rev. B}
	\textbf{\bibinfo{volume}{76}}, \bibinfo{pages}{052203}
	(\bibinfo{year}{2007}).
	
	\bibitem[{\citenamefont{Middleton and Wingreen}(1993)}]{PhysRevLett.71.3198}
	\bibinfo{author}{\bibfnamefont{A.~A.} \bibnamefont{Middleton}}
	\bibnamefont{and} \bibinfo{author}{\bibfnamefont{N.~S.}
		\bibnamefont{Wingreen}}, \bibinfo{journal}{Phys. Rev. Lett.}
	\textbf{\bibinfo{volume}{71}}, \bibinfo{pages}{3198} (\bibinfo{year}{1993}).
	
\end{thebibliography}

\end{document}